\documentclass[journal,10pt,twoside]{IEEEtran}

\usepackage{amsmath,cite,amsfonts,amssymb,psfrag}
\usepackage{amsthm}
\usepackage{graphicx}
\usepackage{epstopdf}
\usepackage{rotating}
\usepackage{dsfont}
\usepackage{color}
\usepackage{tikz}
\usetikzlibrary{plotmarks}
\usetikzlibrary{shapes,arrows,fit,calc,positioning,automata}
\usepackage{pgfplots}
\usetikzlibrary{calc}
\usetikzlibrary{shapes,arrows}
\usetikzlibrary{decorations.markings}
\usetikzlibrary{positioning}
\pgfplotsset{compat=1.10}
\usetikzlibrary{calc}
\usetikzlibrary{shapes,arrows}
\usetikzlibrary{decorations.markings}

\newcommand{\Enc}{\mathsf{Enc}}
\newcommand{\Dec}{\mathsf{Dec}}

\newcommand*\xor{\mathbin{\oplus}}

\makeatletter
\newtheorem*{rep@theorem}{\rep@title}
\newcommand{\newreptheorem}[2]{%
	\newenvironment{rep#1}[1]{%
		\def\rep@title{\Cref{##1}}%
		\begin{rep@theorem}}%
		{\end{rep@theorem}}}
\newcommand*{\textlabel}[2]{%
	\edef\@currentlabel{#1}
	\phantomsection
	#1\label{#2}
}
\makeatother

\makeatletter
\newsavebox\myboxA
\newsavebox\myboxB
\newlength\mylenA

\newcommand*\xoverline[2][0.75]{%
	\sbox{\myboxA}{$\m@th#2$}%
	\setbox\myboxB\null
	\ht\myboxB=\ht\myboxA%
	\dp\myboxB=\dp\myboxA%
	\wd\myboxB=#1\wd\myboxA
	\sbox\myboxB{$\m@th\overline{\copy\myboxB}$}
	\setlength\mylenA{\the\wd\myboxA}
	\addtolength\mylenA{-\the\wd\myboxB}%
	\ifdim\wd\myboxB<\wd\myboxA%
	\rlap{\hskip 0.5\mylenA\usebox\myboxB}{\usebox\myboxA}%
	\else
	\hskip -0.5\mylenA\rlap{\usebox\myboxA}{\hskip 0.5\mylenA\usebox\myboxB}%
	\fi}
\makeatother

\newtheorem{theorem}{Theorem}

\newtheorem{remark}{Remark}

\newtheorem{definition}{Definition}

\begin{document}
\title{Code Constructions for Physical Unclonable Functions and Biometric Secrecy Systems}
\IEEEoverridecommandlockouts

\author{Onur G\"unl\"u,	Onurcan \.{I}\c{s}can, Vladimir Sidorenko,~\IEEEmembership{Member,~IEEE,} and Gerhard Kramer,~\IEEEmembership{Fellow,~IEEE}
	
	\thanks{Manuscript received August 11, 2018; revised January 22, 2019; accepted March 28, 2019. The work of O. G\"unl\"u was supported by the German Research Foundation (DFG) through the Project HoliPUF under the grant KR3517/6-1. V. Sidorenko is on leave from the Institute for Information Transmission Problems, Russian Academy of Science. The work of V. Sidorenko is supported by the Russian Government (Contract No 14.W03.31.0019). The work of G. Kramer was supported by an Alexander von Humboldt Professorship endowed by the German Federal Ministry of Education and Research. Parts of this paper are available online in \cite{OurWZArxiv}. The associate editor coordinating the review of this manuscript and approving it for publication was Dr. Shantanu Rane (\textit{Corresponding Author: Onur G\"unl\"u}).}
	\thanks{O. G\"unl\"u, V. Sidorenko, and G. Kramer are with the Chair of Communications Engineering, Technical University of Munich, 80333 Munich, Germany (e-mail: \{onur.gunlu, vladimir.sidorenko, gerhard.kramer\}@tum.de).
		
	O. \.{I}\c{s}can is with Huawei Technologies Duesseldorf GmbH, 80992 Munich, Germany (email: onurcan.iscan@huawei.com).}
}
\maketitle

\begin{abstract}
The two-terminal key agreement problem with biometric or physical identifiers is considered. Two linear code constructions based on Wyner-Ziv coding are developed. The first construction uses random linear codes and achieves all points of the key-leakage-storage regions of the generated-secret and chosen-secret models. The second construction uses nested polar codes for vector quantization during enrollment and for error correction during reconstruction. Simulations show that the nested polar codes achieve privacy-leakage and storage rates that improve on existing code designs. One proposed code achieves a rate tuple that cannot be achieved by existing methods.
\end{abstract}

\section{Introduction}
\IEEEPARstart{B}{iometric} features like fingerprints can be used to authenticate and identify individuals, and to generate secret keys. Similarly, one can generate secret keys with physical unclonable functions (PUFs) that are used as sources of randomness. For example, fine variations of ring oscillator (RO) outputs and the start-up behavior of static random access memories (SRAM) can serve as PUFs \cite{bizimtemperature}. Fingerprints and PUFs are identifiers with high entropy and reliable outputs \cite{IgnaTrans,GassendThesis}, and one can consider them as physical ``one-way functions" that are easy to compute and difficult to invert \cite{PappuThesis}. 

There are several requirements that a PUF-based key agreement method should fulfill. First, the method should not leak information about the secret key (no \textit{secrecy leakage}). Second, the method should leak little information about the identifier (limited \textit{privacy leakage}). For example, in some applications the same identifier is enrolled multiple times. If the eavesdropper can extract more information about the identifier each time it is enrolled, then the eavesdropper might be able to learn the secret key of some enrollments. Third, one should limit the \textit{storage} rate because storage can be expensive and limited, e.g., for internet-of-things (IoT) devices. 
\vspace{-0.35cm}
\subsection{Related Work and on Basic PUF Models}
There are two common models for key agreement: the \textit{generated-secret (GS)} and the \textit{chosen-secret (CS) models}. For the GS model, an encoder extracts a secret key from an identifier measurement, while for the CS model a secret key that is independent of the identifier measurements is given to the encoder by a trusted entity. For the key-agreement model introduced in \cite{AhlswedeCsiz} and \cite{Maurer}, two terminals observe dependent random variables and have access to an authenticated, public, one-way communication link; an eavesdropper observes the public messages, called \textit{helper data}. The GS model is treated in \cite[Thm. 2.6]{csiszarnarayan} as a special case of a more general key agreement problem with eavesdropper side information and a helper. However, \cite{AhlswedeCsiz,Maurer,csiszarnarayan} do not consider the privacy leakage. The regions of achievable secret-key vs. privacy-leakage (key-leakage) rates for the GS and CS models are given in \cite{IgnaTrans,LaiTrans}. The storage rates for general (non-negligible) secrecy-leakage levels are analyzed in \cite{storage}, while the rate regions with multiple encoder and decoder measurements of a hidden source are treated in \cite{bizimTIFSMultipleMeasurement}. 

The above papers consider identifier measurements that are independent and identically distributed (i.i.d.) according to a probability distribution with a discrete alphabet. We remark that raw identifier outputs usually have memory but there are transform coding algorithms \cite[pp. 76]{Transformbio},\cite{OurEntropy} that can extract almost i.i.d. and uniformly distributed bits from identifier outputs.

\subsection{Other Models}
There are many other key-agreement models. For instance, key agreement and device authentication with an eavesdropper that has access to a sequence correlated with the identifier outputs has been studied in \cite{csiszarnarayan,Khisti,Blochpaper,OurKittipongTIFS}. This model with eavesdropper side information may be unrealistic for the applications we consider because many physical and biometric identifiers are used for \textit{on-demand} key reconstruction. This means that the attack should be performed during execution, and an invasive attack applied to obtain a correlated sequence permanently changes the identifier output \cite{GassendThesis}.

A closely related problem to the key agreement problem is Wyner's wiretap channel \cite{WynerWTC}, for which code constructions are studied in, e.g., \cite{WTCpolarVardy,KliewerWTC,OzanWTC}. The main aim in this problem is to hide a transmitted message from the eavesdropper that observes a channel output correlated with the observation of a legitimate receiver. 

\subsection{Summary of Contributions}
We propose code constructions for the key agreement models of \cite{IgnaTrans,LaiTrans,bizimTIFSMultipleMeasurement} and illustrate that they are asymptotically optimal and improve on all existing methods. The code constructions are based on Wyner-Ziv (WZ) coding \cite{WZratedistortion}. A summary of the main contributions is as follows.
\begin{itemize}	
	\item We describe two WZ-coding constructions for binary symmetric sources and binary symmetric channels (BSCs). Such models are often used for physical identifiers such as RO PUFs \cite{OurEntropy} and SRAM PUFs \cite{maes2009soft}. The first construction is based on \cite{lossysourcecoding} and achieves all points of the key-leakage-storage regions of the GS and CS models. The novelty is that we propose additional steps to specify a secret key and show that the construction is optimal.     
	\item The second construction uses nested polar codes. We design and simulate our polar codes for standard parameters for SRAM PUFs under ideal environmental conditions, and for RO PUFS under varying environmental conditions. The target block-error probability is $P_B=10^{-6}$ and the target secret-key size is 128 bits. One of the codes achieves key-leakage-storage rates that cannot be achieved by existing methods.
	\item In Appendix~\ref{app:strongsecrecy}, we consider strong secrecy.
	\item In Appendix~\ref{app:hiddenidentifiers}, we consider a hidden identifier source whose noisy measurements via BSCs are observed at the encoder and decoder. The WZ-coding construction is shown to be optimal also for such identifiers.
\end{itemize}

\subsection{Organization}
This paper is organized as follows. In Section~\ref{sec:problem_settingandcode}, we describe the GS and CS models, the WZ problem, and give their rate regions. We show that existing methods are suboptimal even after applying improvements described in Section~\ref{sec:CompareMethods}. Section~\ref{sec:firstWZconstruction} describes a random linear code construction based on WZ-coding. Section~\ref{sec:codeproposal} describes a nested polar code design for the GS model and illustrates that it improves on existing code designs.

\subsection{Notation}
Upper case letters represent random variables and lower case letters their realizations. A superscript denotes a string of variables, e.g., $\displaystyle X^n\!=\!X_1\ldots X_i\ldots X_n$, and a subscript denotes the position of a variable in a string. A random variable $\displaystyle X$ has probability distribution $\displaystyle P_X$. Calligraphic letters such as $\displaystyle \mathcal{X}$ denote sets, and set sizes are written as $\displaystyle |\mathcal{X}|$. Bold letters such as $\mathbf{H}$ represent matrices. $\Enc(\cdot)$ is an encoder mapping and $\Dec(\cdot)$ is a decoder mapping. $H_b(x)=-x\log x- (1-x)\log (1-x)$ is the binary entropy function, where we take logarithms to the base $2$. The $*$-operator is defined as $\displaystyle p*x = p(1-x)+(1-p)x$. The operator $\xor$ represents the element-wise modulo-2 summation. A BSC with crossover probability $p$ is denoted by BSC($p$). $X^n\sim\text{Bern}^n(\alpha)$ is an i.i.d. binary sequence of random variables with $\Pr[X_i=1]=\alpha$ for $i=1,2,\ldots,n$. ${\mathbf{H}}^T$ represents the transpose of $\mathbf{H}$. A linear error-correction code with parameters $(n,k)$ has block length $n$ and dimension $k$.

\section{Problem Formulations}\label{sec:problem_settingandcode}
\subsection{Generated-secret and Chosen-secret Models}\label{subsec:GSCSmodel} 
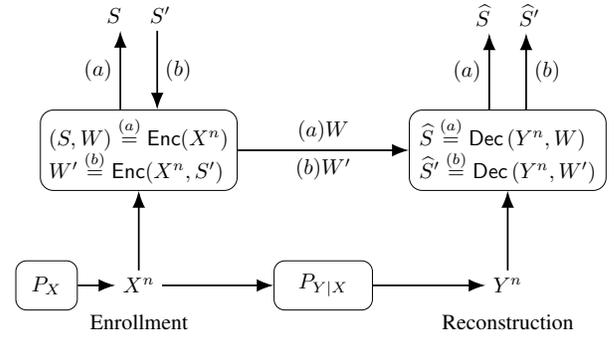
\begin{figure}
			\centering
	\resizebox{0.91\linewidth}{!}{

		\begin{tikzpicture}
		\node (so) at (-1.5,-2.2) [draw,rounded corners = 5pt, minimum width=1.0cm,minimum height=0.8cm, align=left] {$P_X$};
		\node (a) at (0,0) [draw,rounded corners = 6pt, minimum width=3.2cm,minimum height=1.2cm, align=left] {$
			(S,W) \overset{(a)}{=} \Enc(X^n)$\\ $W'\overset{(b)}{=}\Enc(X^n,S')$};
		\node (c) at (3,-2.2) [draw,rounded corners = 5pt, minimum width=1.6cm,minimum height=0.8cm, align=left] {$P_{Y|X}$};
		\node (b) at (6,0) [draw,rounded corners = 6pt, minimum width=3.2cm,minimum height=1.2cm, align=left] {$\widehat{S} \overset{(a)}{=} \Dec\left(Y^n,W\right)$\\$\widehat{S}' \overset{(b)}{=} \Dec\left(Y^n,W'\right)$};
		\draw[decoration={markings,mark=at position 1 with {\arrow[scale=1.5]{latex}}},
		postaction={decorate}, thick, shorten >=1.4pt] (a.east) -- (b.west) node [midway, above] {$(a) W$} node [midway, below] {$(b) W'$};;
		\node (a1) [below of = a, node distance = 2.2cm] {$X^n$};
		\node (b1) [below of = b, node distance = 2.2cm] {$Y^n$};
		\node (k9) [below of = a1, node distance = 0.6cm] {Enrollment};
		\node (k19) [below of = b1, node distance = 0.6cm] {Reconstruction};
		\draw[decoration={markings,mark=at position 1 with {\arrow[scale=1.5]{latex}}},
		postaction={decorate}, thick, shorten >=1.4pt] (so.east) -- (a1.west);
		\draw[decoration={markings,mark=at position 1 with {\arrow[scale=1.5]{latex}}},
		postaction={decorate}, thick, shorten >=1.4pt] (a1.north) -- (a.south);
		\draw[decoration={markings,mark=at position 1 with {\arrow[scale=1.5]{latex}}},
		postaction={decorate}, thick, shorten >=1.4pt] (a1.east) -- (c.west);
		\draw[decoration={markings,mark=at position 1 with {\arrow[scale=1.5]{latex}}},
		postaction={decorate}, thick, shorten >=1.4pt] (c.east) -- (b1.west);
		\draw[decoration={markings,mark=at position 1 with {\arrow[scale=1.5]{latex}}},
		postaction={decorate}, thick, shorten >=1.4pt] (b1.north) -- (b.south);
		\node (a2) [above of = a, node distance = 2.2cm] {$S\quad\,\, S'$};
		\node (b2) [above of = b, node distance = 2.2cm] {$\widehat{S}\quad\,\,\widehat{S}'$};
		\draw[decoration={markings,mark=at position 1 with {\arrow[scale=1.5]{latex}}},
		postaction={decorate}, thick, shorten >=1.4pt] ($(b.north)-(0.3,0)$) -- ($(b2.south)-(0.3,0)$) node [midway, left] {$(a)$};
				\draw[decoration={markings,mark=at position 1 with {\arrow[scale=1.5]{latex}}},
				postaction={decorate}, thick, shorten >=1.4pt] ($(b.north)+(0.3,0)$)-- ($(b2.south)+(0.3,0)$) node [midway, right] {$(b)$};
		\draw[decoration={markings,mark=at position 1 with {\arrow[scale=1.5]{latex}}},
		postaction={decorate}, thick, shorten >=1.4pt] ($(a.north)-(0.3,0)$)-- ($(a2.south)-(0.3,0)$) node [midway, left] {$(a)$};
		\draw[decoration={markings,mark=at position 1 with {\arrow[scale=1.5]{latex}}},
		postaction={decorate}, thick, shorten >=1.4pt]  ($(a2.south)+(0.3,0)$)-- ($(a.north)+(0.3,0)$) node [midway, right] {$(b)$};
		\end{tikzpicture}
	}
	\caption{The $(a)$ GS and $(b)$ CS models.}\label{fig:problemsetup}
\end{figure}

Consider the GS model in Fig.~\ref{fig:problemsetup}$(a)$, where a secret key is generated from a biometric or physical source. The source, measurement, secret key, and storage alphabets $\mathcal{X}$, $\mathcal{Y}$, $\mathcal{S}$, and $\mathcal{W}$ are finite sets. During enrollment, the encoder observes an i.i.d. sequence $X^n$, generated by the identifier (source) according to some $P_X$, and computes a secret key $S$ and public helper data $W$ as $\displaystyle (S,W)\,{=}\,{\Enc}(X^n)$. During reconstruction, the decoder observes a noisy source measurement $Y^n$ of $X^n$ through a memoryless channel $P_{Y|X}$ together with the helper data $W$. The decoder estimates the secret key as $\displaystyle \widehat{S}\,{=}\,{\Dec}(Y^n\!,W)$. Similarly, Fig.~\ref{fig:problemsetup}$(b)$ shows the CS model, where a secret key $S'\in\mathcal{S}$ that is independent of $(X^n,Y^n)$ is embedded into the helper data as $W' = \Enc(X^n,S')$. The decoder for the CS model estimates the secret key as $\widehat{S}'=\Dec(Y^n,W')$.

\begin{definition}\label{def:achievabilityGSCS}
	 A key-leakage-storage tuple $(R_s,R_\ell,R_w)$ is \emph{achievable} for the GS model if, given any $\epsilon>0$, there is some $n\!\geq\!1$, an encoder, and a decoder such that $R_s=\frac{\log|\mathcal{S}|}{n}$ and
	\begin{align}
	&\Pr[\widehat{S} \neq S] \leq \epsilon&&\quad (\text{reliability})\label{eq:reliability_constraint}\\
	&\frac{1}{n}I(S;W) \leq \epsilon&&\quad(\text{weak secrecy})\label{eq:secrecyleakage_constraint}\\
	&\frac{1}{n}H(S)\geq R_s-\epsilon&&\quad(\text{key uniformity}) \label{eq:uniformity_constraint}\\
	&\frac{1}{n}\log\big|\mathcal{W}\big| \leq R_w+\epsilon&&\quad(\text{storage})\label{eq:storage_constraint}\\
	&\frac{1}{n}I(X^n;W) \leq R_\ell+\epsilon&&\quad(\text{privacy}).\label{eq:leakage_constraint}
	\end{align}
	Similarly, a tuple $(R_s,R_\ell,R_w)$ is \emph{achievable} for the CS model if, given any $\epsilon>0$, there is some $n\!\geq\!1$, an encoder, and a decoder such that $R_s=\frac{\log|\mathcal{S}|}{n}$ and (\ref{eq:reliability_constraint})-(\ref{eq:leakage_constraint}) are satisfied when $S$ and $W$ are replaced by, respectively, $S'$ and $W'$.
	 
    The \emph{key-leakage-storage} regions $\mathcal{R}_{\text{gs}}$ and $\mathcal{R}_{\text{cs}}$ for the GS and CS models, respectively, are the closures of the sets of achievable tuples for the corresponding models.\hfill $\lozenge$
\end{definition}

\begin{theorem}[\hspace{1sp}\cite{IgnaTrans}]\label{theo:secrecyregions}
The key-leakage-storage regions for the GS and CS models, respectively, are 
\begin{align}
\mathcal{R}_{\text{gs}}\! =\! &\bigcup_{P_{U|X}}\!\Big\{\left(R_s,R_\ell,R_w\right)\!\colon\!\nonumber\\
&0 \leq R_s\leq I(U;Y),\nonumber\\
&R_\ell\geq I(U;X)-I(U;Y),\nonumber\\
&R_w\geq I(U;X)-I(U;Y)\Big\} \text{,}\label{eq:regionGS}
\end{align}
\vspace*{-0.2cm}
\begin{align}
	\mathcal{R}_{\text{cs}}\! =\! &\bigcup_{P_{U|X}}\!\Big\{\left(R_s,R_\ell,R_w\right)\!\colon\! \nonumber\\
	&0 \leq R_s\leq I(U;Y),\nonumber\\
	&R_\ell\geq I(U;X)-I(U;Y),\nonumber\\
	&R_w\geq I(U;X)\Big\} \label{eq:regionCS}
\end{align}
where $U-X-Y$ forms a Markov chain. These regions are convex sets. The alphabet $\mathcal{U}$ of the auxiliary random variable $U$ can be limited to have size $\displaystyle |\mathcal{U}|\!\leq\!|\mathcal{X}|+1$.
\end{theorem}

For example, suppose $X^n\sim \text{Bern}^n(\frac{1}{2})$ and the channel $P_{Y|X}$ is a BSC$(p_A)$, where $p_A\in[0, 0.5]$. The key-leakage-storage region of the GS model for this case is \cite{IgnaTrans}
\begin{align}
\mathcal{R}_{\text{gs,bin}}\! =\! &\bigcup_{q\in[0,0.5]}\!\Big\{\left(R_s,R_\ell,R_w\right)\!\colon\!\nonumber\\
&0\leq R_s\leq 1- H_b(q*p_A),\nonumber\\
&R_\ell\geq H_b(q*p_A)- H_b(q),\nonumber\\
&R_w\geq H_b(q*p_A)- H_b(q)\Big\}\label{eq:BSCRegionGS}.
\end{align}

\subsection{Wyner-Ziv Problem} \label{subsec:systemmodelWZ}
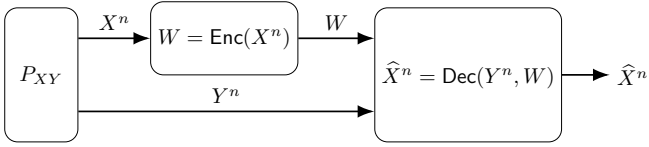
\begin{figure}
 	\resizebox{1.0\linewidth}{!}{
 		\centering
 		\begin{tikzpicture}
 		
 		\node (a) at (0,0) [draw,rounded corners = 6pt, minimum width=1.2cm,minimum height=2.2cm, align=left] {$P_{XY}$};
 		\node (b) at (7,0) [draw,rounded corners = 6pt, minimum width=2.2cm,minimum height=2.2cm, align=left] {$\widehat{X}^n=\Dec(Y^n,W)$};
 		\node (d) at (3,0.6) [draw,rounded corners = 6pt, minimum width=1.2cm,minimum height=1.2cm, align=left] {$W=\Enc(X^n)$};
 		
 		\draw[decoration={markings,mark=at position 1 with {\arrow[scale=1.5]{latex}}},
 		postaction={decorate}, thick, shorten >=1.4pt] ($(a.east)+(0,0.6)$) -- ($(d.west)+(0,0)$) node [midway, above] {$X^n$};;
 		\draw[decoration={markings,mark=at position 1 with {\arrow[scale=1.5]{latex}}},
 		postaction={decorate}, thick, shorten >=1.4pt] ($(d.east)+(0,0)$) -- ($(b.west)+(0,0.6)$) node [midway, above] {$W$};;
 		\draw[decoration={markings,mark=at position 1 with {\arrow[scale=1.5]{latex}}},
 		postaction={decorate}, thick, shorten >=1.4pt] ($(a.east)-(0,0.6)$) -- ($(b.west)-(0,0.6)$) node [midway, above] {$Y^n$};;
 		\node (b2) [right of = b, node distance = 2.7cm] {$\widehat{X}^n$};
 		\draw[decoration={markings,mark=at position 1 with {\arrow[scale=1.5]{latex}}},
 		postaction={decorate}, thick, shorten >=1.4pt] (b.east) -- (b2.west);
 		\end{tikzpicture}
 	}
 	\caption{The WZ problem.}\label{fig:WZsetup}
 \end{figure}

Consider two dependent random variables $X$ and $Y$ with joint distribution $P_{XY}$. Fig.~\ref{fig:WZsetup} depicts the WZ problem. The source, side information, and message alphabets $\mathcal{X}$, $\mathcal{Y}$, and $\mathcal{W}$ are finite sets. An encoder that observes $X^n$ generates the message $W$. The decoder observes $Y^n$ and $W$ and puts out a quantized version $\widehat{X}^n$ of $X^n$. Define the average distortion between $X^n$ and the reconstructed sequence $\widehat{X}^n$ as 
\begin{align}
\frac{1}{n}\sum_{i=1}^nE[d(X_i, \widehat{X}_i(Y^n,W))]
\end{align}  
where $d(x,\hat{x})$ is a bounded distortion function and $\widehat{X}_i(y^n,w)$ is a reconstruction function.

\begin{definition}\label{def:WZdefinition}
	 A rate-distortion pair $(R_w,D)$ is \emph{achievable} if, given any $\epsilon>0$, there is some $n\geq 1$, an encoder, and a decoder that satisfy (\ref{eq:storage_constraint}) and
	\begin{align}
	\frac{1}{n}\sum_{i=1}^nE[d(X_i, \widehat{X}_i(Y^n,W))]\leq D+\epsilon \label{eq:WZdistortion}.
	\end{align}
	The WZ rate-distortion region $\mathcal{R}_{\text{WZ}}$ is the closure of the set of achievable rate-distortion pairs.\hfill $\lozenge$
\end{definition}

\begin{theorem}[\hspace{1sp}\cite{WZratedistortion}]\label{theo:WZ}
	The WZ rate-distortion region is 
	\begin{align}
	\mathcal{R}_{\text{WZ}}\! =\! &\bigcup_{P_{U|X}}\bigcup_{\widehat{X}(Y,U)}\!\Big\{\left(R_w,D\right)\!\colon\nonumber\\
	&R_w\geq I(U;X)-I(U;Y),\nonumber\\
	&D\geq E[d(X, \widehat{X}(Y,U))]\Big\} \label{eq:WZrateregion}
	\end{align}
	where $U-X-Y$ forms a Markov chain and $\widehat{X}(Y,U)$ is a reconstruction function. One can limit the alphabet $\mathcal{U}$ of the auxiliary random variable $U$ to have size $\displaystyle |\mathcal{U}|\!\leq\!|\mathcal{X}|+1$. The region $\mathcal{R}_{\text{WZ}}$ is convex.
\end{theorem}

\section{Prior Art and Comparisons}\label{sec:CompareMethods}
There are several existing code constructions proposed for the GS and CS models. We consider the three best methods: the fuzzy-commitment scheme (FCS) \cite{FuzzyCommitment} for the CS model, the code-offset fuzzy extractor (COFE) \cite{Dodis2008fuzzy}, and the polar code construction in \cite{IgnaPolar} for the GS model. We show that these constructions are suboptimal in terms of the privacy-leakage and storage rates.

The binary Golay code is used in \cite{IgnaTrans} as a vector quantizer (VQ) in combination with Slepian-Wolf (SW) codes \cite{SW} to illustrate that the key vs. storage (or key vs. leakage) rate ratio can be increased via quantization. This observation motivates using a VQ to improve the performance of previous constructions. In Sections~\ref{sec:firstWZconstruction} and \ref{sec:codeproposal}, we apply VQ by using WZ coding to decrease storage rates, as suggested in \cite[Remark 4.5]{Blochbook}.

During enrollment with the FCS, an encoder takes a uniformly distributed secret key $S'$ as input to generate a codeword $C^n$. The codeword and the binary source output $X^n$ are summed modulo-2, and the sum is stored as helper data $W'$. During reconstruction, $W'$ and a binary sequence $Y^n$ are summed modulo-2 and this sum is used by a decoder to estimate $S'$. Similar steps are applied in the COFE, except that the secret key is a hashed version of $X^n$. The FCS achieves one optimal point in the key-leakage region, namely the point with the maximum secret-key rate $R_s^*=I(X;Y)$ and the privacy-leakage rate $R_\ell^* =H(X|Y)$ \cite{IgnaFuzzy}. Similarly, the COFE achieves the same boundary point in the key-leakage region. This is, however, the only boundary point that these methods can achieve.

We can improve both methods by adding a VQ step: instead of $X^n$ we use its quantized version $X_q^n$ during enrollment. This asymptotically corresponds to summing the original helper data and an independent random variable $J^n\sim\text{Bern}^n(q)$ such that $W''=X^n\xor C^n\xor J^n$ is the new helper data so that we create a virtual channel $P_{Y|X\xor J}$ and apply the FCS or COFE to this virtual channel. The modified FCS and COFE can achieve all points of the key-leakage region if we take a union of all rate pairs achieved over all $q\in[0, 0.5]$. However, the helper data has a length of $n$ bits for both methods, and the resulting storage rate of $1$ bit/symbol is not necessarily optimal.

The polar code construction in \cite{IgnaPolar} requires less storage rate than the FCS and COFE. However, this approach improves only the storage rate and cannot achieve all points of the key-leakage-storage region. Furthermore, in \cite{IgnaPolar} some code designs assume that there is a ``private" key shared only between the encoder and decoder, which is not realistic since a private key requires hardware protection against invasive attacks. If such a protection is possible, then there is no need to use an on-demand key reconstruction method like a PUF. 

The existing methods cannot, therefore, achieve all points of the key-leakage-storage region for a BSC, unlike the WZ-coding constructions we describe in Sections~\ref{sec:firstWZconstruction} and \ref{sec:codeproposal}. {There is another code construction in \cite{RemiPolarConstruction} that is optimal for the GS and CS models. However, its hardware complexity seems to be high and it has unrealistic requirements similar to the private key required for some code designs in \cite{IgnaPolar}; see Remark~\ref{rem:RemiPolarNotPractical} below.} 

In previous works such as \cite{Pufky}, only the secret-key rates of the proposed codes are compared because the sum of the secret-key and storage (or privacy-leakage) rates is fixed. This constraint means that increasing the key vs. storage (or key vs. leakage) rate ratio is equivalent to increasing the key rate. Instead, our code constructions are more flexible in terms of achievable rate tuples. We will use the key vs. storage rate ratio as a metric to control the storage and privacy leakage in our code designs.

\section{First WZ-coding Construction}\label{sec:firstWZconstruction}
Consider the lossy source coding construction proposed in \cite[Section IV]{lossysourcecoding} that achieves the boundary points of the WZ rate-distortion region by using linear codes. We use this construction to achieve the boundary points of $\mathcal{R}_{\text{gs}}$ and $\mathcal{R}_{\text{cs}}$ for a binary uniform identifier source $P_X$ and a BSC $P_{Y|X}$ with crossover probability $p_A$. 

\subsection{Review of a WZ-coding Construction}
Consider the WZ problem depicted in Fig.~\ref{fig:WZsetup}.

\emph{Code Construction}: Choose uniformly at random the full-rank parity-check matrices $\mathbf{H}_1$, $\mathbf{H}_2$, and $\mathbf{H}$ as
\begin{align}
\mathbf{H} = \begin{bmatrix}
\mathbf{H}_1\\[0.3em]
\mathbf{H}_2
\end{bmatrix}\label{eq:Hdef}
\end{align}
where $\mathbf{H}_1$ with dimensions $m_1\times n$ defines a binary $(n,n-m_1)$ linear code $\mathcal{C}_1$ and $\mathbf{H}_2$ with dimensions $m_2\times n$ defines another binary $(n,n\!-\!m_2)$ linear code $\mathcal{C}_2$. The $(n,n\!-\!m_1\!-\!m_2)$ code $\mathcal{C}$ defined by $\mathbf{H}$ in (\ref{eq:Hdef}) is thus a subcode of $\mathcal{C}_1$ such that $\mathcal{C}_1$ is partitioned into $2^{m_2}$ cosets of $\mathcal{C}$. For some distortion $q\in[0,0.5]$ and $\delta>0$, impose the conditions
\begin{align}
&\frac{m_1}{n} = H_b(q)-\delta\label{eq:constraintonm1}\\ 
&\frac{m_1+m_2}{n} = H_b(q*p_A)+\delta\label{eq:constraintonm2}.
\end{align}   

\emph{Encoding}: A VQ quantizes the source output $X^n$ to the closest codeword $X_{q}^n$ in $\mathcal{C}_1$ in Hamming metric. If there are two or more codewords with the minimum Hamming distance, the VQ chooses one of them. Define the error sequence 
\begin{align}
E_{q}^n = X^n\xor X_{q}^n \label{eq:minhammdist}
\end{align}
which resembles an i.i.d. sequence $\sim\text{Bern}^n(q)$ when $n\rightarrow \infty$ due to uniformity of $X^n$ and the linearity of $\mathcal{C}_1$ \cite{lossysourcecoding}. 

We publicly store the message 
\begin{align}
W = X_{q}^{n}\mathbf{H}_2^T\label{eq:whattostore}
\end{align}
which corresponds to a coset of $\mathcal{C}$. 

\emph{Decoding}: The decoder sees $Y^n = X^n\xor Z^n$, where $Z^n$ is independent of $X^n$ and $Z^n\sim\text{Bern}^n(p_A)$. The error sequence $E_{q}^n$ and the noise sequence $Z^n$ are independent. Furthermore, $E_{q}^n$ asymptotically resembles an i.i.d. sequence $\sim\text{Bern}^n(q)$ when $n\rightarrow \infty$, as discussed above. Therefore, when $n\rightarrow\infty$, the sequence $E_{q}^n\xor Z^n$, which corresponds to the noise sequence of the equivalent channel $\displaystyle P_{Y^n|X_q^n}$, is distributed according to $\text{Bern}^n(q*p_A)$ since the equivalent channel is a concatenation of two BSCs. One can thus reconstruct $X_{q}^n$ with high probability when $n\rightarrow\infty$ by using the syndrome decoder $f_{\mathcal{C}}(\cdot)$ of the code $\mathcal{C}$ as follows
\begin{align}
\widehat{X}_{q}^n& = Y^n\xor {f_{\mathcal{C}}([0,\, W]\xor Y^{n}\mathbf{H}^T)}\nonumber\\ 
&\overset{(a)}{=} Y^n\xor {f_{\mathcal{C}}(X_{q}^{n}\mathbf{H}^T\xor Y^{n}\mathbf{H}^T)}\nonumber\\
&\overset{(b)}{=}  (X_{q}^n\xor E_{q}^n\xor Z^n)\xor {f_{\mathcal{C}}((E_{q}^{n}\xor Z^{n})\mathbf{H}^T)}\nonumber\\
&\overset{(c)}{=} (X_{q}^n\xor E_{q}^n\xor Z^n)\xor (E_{q}^n\xor Z^n)\nonumber\\
&=X_{q}^n\label{eq:estimateXqn}
\end{align}
where $(a)$ follows by (\ref{eq:whattostore}) and because $X_{q}^n$ is a codeword of $\mathcal{C}_1$, $(b)$ follows by (\ref{eq:minhammdist}), and $(c)$ follows with high probability because, asymptotically, $E_{q}^n\xor Z^n\sim \text{Bern}^n(q*p_A)$ so that the syndrome decoder $f_{\mathcal{C}}(\cdot)$ determines the noise sequence $E_{q}^n\xor Z^n$. This is because the constraint in (\ref{eq:constraintonm2}) indicates that the code rate of $\mathcal{C}$ is below the capacity of the BSC$(q*p_A)$.

\subsection{Key Agreement}\label{subsec:SKWZfirstconstruct}
\begin{figure}
	\centering
	\resizebox{1.01\linewidth}{!}{	
		\begin{tikzpicture}
		\node (so) at (-1.5,-2.2) [draw,rounded corners = 5pt, minimum width=1.0cm,minimum height=0.8cm, align=left] {$P_X$};
		\node (a) at (0,0) [draw,rounded corners = 6pt, minimum width=3.2cm,minimum height=1.2cm, align=left] {$
			X_{q}^n = \text{VQ}\left(\mathbf{H}_1,X^n\right)$\\[0.1cm]
			$W= X_{q}^{n}\mathbf{H}_2^{T},\;\; S\!=\!\Dec_{\mathcal{C}}(X_q^n)$\\[0.05cm]   
			$W'\overset{(b)}{=}\left[W,\,S\xor S'\right]$};
		\node (c) at (3,-2.2) [draw,rounded corners = 5pt, minimum width=1.6cm,minimum height=0.8cm, align=left] {$P_{Y|X}$};
		\node (b) at (6,0) [draw,rounded corners = 6pt, minimum width=3.2cm,minimum height=1.2cm, align=left] 
		{$\widehat{X}_q^n=Y^n\xor {f_{\mathcal{C}}([0,\, W]\xor Y^{n}\mathbf{H}^T)}$\\[0.1cm]
			$\widehat{S}\!=\!\Dec_{\mathcal{C}}(\widehat{X}_q^n)$\\[0.1cm]
			$\widehat{S}'\overset{(b)}{=}\widehat{S}\xor (S\xor S')$};
		\draw[decoration={markings,mark=at position 1 with {\arrow[scale=1.5]{latex}}},
		postaction={decorate}, thick, shorten >=1.4pt] (a.east) -- (b.west) node [midway, above] {$(a) W$} node [midway, below] {$(b) W'$};
		\node (a1) [below of = a, node distance = 2.2cm] {$X^n$};
		\node (b1) [below of = b, node distance = 2.2cm] {$Y^n$};
		\node (k9) [below of = a1, node distance = 0.6cm] {Enrollment};
		\node (k19) [below of = b1, node distance = 0.6cm] {Reconstruction};
		\draw[decoration={markings,mark=at position 1 with {\arrow[scale=1.5]{latex}}},
		postaction={decorate}, thick, shorten >=1.4pt] (so.east) -- (a1.west);
		\draw[decoration={markings,mark=at position 1 with {\arrow[scale=1.5]{latex}}},
		postaction={decorate}, thick, shorten >=1.4pt] (a1.north) -- (a.south);
		\draw[decoration={markings,mark=at position 1 with {\arrow[scale=1.5]{latex}}},
		postaction={decorate}, thick, shorten >=1.4pt] (a1.east) -- (c.west);
		\draw[decoration={markings,mark=at position 1 with {\arrow[scale=1.5]{latex}}},
		postaction={decorate}, thick, shorten >=1.4pt] (c.east) -- (b1.west);
		\draw[decoration={markings,mark=at position 1 with {\arrow[scale=1.5]{latex}}},
		postaction={decorate}, thick, shorten >=1.4pt] (b1.north) -- (b.south);
		\node (a2) [above of = a, node distance = 2.2cm] {$S\quad\,\, S'$};
		\node (b2) [above of = b, node distance = 2.2cm] {$\widehat{S}\quad\,\,\widehat{S}'$};
		\draw[decoration={markings,mark=at position 1 with {\arrow[scale=1.5]{latex}}},
		postaction={decorate}, thick, shorten >=1.4pt] ($(b.north)-(0.3,0)$) -- ($(b2.south)-(0.3,0)$) node [midway, left] {$(a)$};
		\draw[decoration={markings,mark=at position 1 with {\arrow[scale=1.5]{latex}}},
		postaction={decorate}, thick, shorten >=1.4pt] ($(b.north)+(0.3,0)$) -- ($(b2.south)+(0.3,0)$) node [midway, right] {$(b)$};
		\draw[decoration={markings,mark=at position 1 with {\arrow[scale=1.5]{latex}}},
		postaction={decorate}, thick, shorten >=1.4pt] ($(a.north)-(0.3,0)$)-- ($(a2.south)-(0.3,0)$) node [midway, left] {$(a)$};
		\draw[decoration={markings,mark=at position 1 with {\arrow[scale=1.5]{latex}}},
		postaction={decorate}, thick, shorten >=1.4pt]  ($(a2.south)+(0.3,0)$)-- ($(a.north)+(0.3,0)$) node [midway, right] {$(b)$};
		\end{tikzpicture}
	}
	\caption{First WZ-coding construction for the $(a)$ GS and $(b)$ CS models, where VQ represents the vector quantization and $\Dec_{\mathcal{C}}$ represents the demapping operation between a codeword of the code $\mathcal{C}$ and the corresponding information sequence.}\label{fig:lossysourcebasedsetup}
\end{figure}
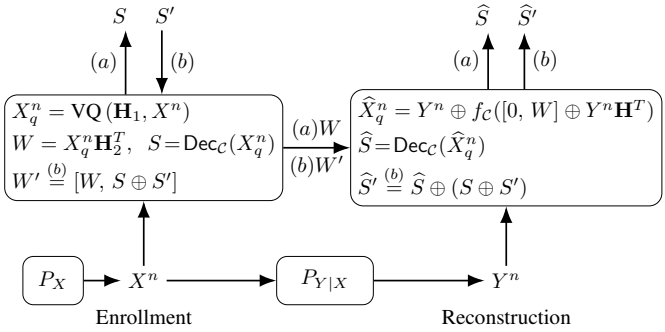

We propose additional steps to agree on a secret key with a negligible secrecy-leakage rate. Figs.~\ref{fig:lossysourcebasedsetup}$(a)$ and \ref{fig:lossysourcebasedsetup}$(b)$ plot the proposed code construction, respectively, for the GS and CS models. 

\emph{Enrollment}: After obtaining the helper data $W$ as in (\ref{eq:whattostore}), we sum modulo-2 the bit sequence that is in the coset $W$ and that has the minimum Hamming weight with $X_q^n$ to obtain a codeword $X_c^n$ of $\mathcal{C}$. Then, we assign the information sequence that is encoded to the codeword $X_c^n$ as the secret key $S$ such that $X_c^n=S\mathbf{G}$, where $\mathbf{G}$ is the generator matrix of $\mathcal{C}$. The secret key has length $n-m_1-m_2$ bits. We denote this operation as $\Dec_{\mathcal{C}}(\cdot)$. 

Consider the secrecy-leakage rate for the GS model:
\begin{align}
&\lim_{n\rightarrow\infty}\frac{1}{n}I(S;W)\! =\! \lim_{n\rightarrow\infty}\frac{1}{n}\!\big(H(S)\!+\!H(W)\!-\!H(W,S)\big)\nonumber\\
&\overset{(a)}{\leq}\lim_{n\rightarrow\infty} \frac{1}{n}\big(\log|\mathcal{S}|+\log|\mathcal{W}|-H(W,S, X_q^n)\big)\nonumber\\
&\leq\lim_{n\rightarrow\infty} \frac{1}{n}\big((n-m_1-m_2)+m_2-H(X_q^n)\big)\nonumber\\
&\overset{(b)}{\leq}\lim_{n\rightarrow\infty} \frac{1}{n}\big(n-m_1-(n-m_1-n\delta_n)\big)=0\label{eq:secrecyleakconstruction}
\end{align}
where $(a)$ follows because $(W,S)$ determines $X_q^n$ and $(b)$ follows with high probability for some $\delta_n$ such that $\lim_{n\rightarrow\infty}\delta_n=0$ due to the translation invariance of the linear code $\mathcal{C}_1$ and the uniformity of $X^n$ (see also the discussions in \cite[Section I]{GuruswamiListDec}).

For the CS model shown in Fig.~\ref{fig:lossysourcebasedsetup}$(b)$, we have access to an embedded (chosen) secret key $S'$ that is independent of $(X^n,Y^n)$ and such that $\mathcal{S}=\mathcal{S}'$. We store the helper data $W' = [W,\, S\xor S']$. The secrecy-leakage rate for the CS model is 
\begin{align}
&\lim_{n\rightarrow\infty}\frac{1}{n}I(S';W')= \lim_{n\rightarrow\infty}\frac{1}{n}I(S';W,S\xor S')\nonumber\\
&\overset{(a)}{=}\lim_{n\rightarrow\infty}\frac{1}{n}\Big(H(S')+H(W,S\xor S')-H(W,S)-H(S')\Big)\nonumber\\
&\leq\lim_{n\rightarrow\infty}\frac{1}{n}\Big(H(W)+H(S\xor S')-H(W,S)\Big)\nonumber\\
&\overset{(b)}{\leq}\lim_{n\rightarrow\infty} \frac{1}{n}\big(\log|\mathcal{W}|+\log|\mathcal{S}|-H(W,S, X_q^n)\big)\nonumber\\
&\overset{(c)}{\leq}\!\lim_{n\rightarrow\infty}\frac{1}{n}\!\big(m_2+(n\!-\!m_1\!-\!m_2)-(n\!-\!m_1\!-\!n\delta_n)\big)\!=\!0
\label{eq:secrecyleakconstructionCS}
\end{align}
where $(a)$ follows because $S'$ is independent of $(W,S)$, $(b)$ follows because $\mathcal{S}=\mathcal{S}'$ and $(W,S)$ determines $X_q^n$, and $(c)$ follows with high probability for some $\delta_n$ such that $\lim_{n\rightarrow\infty}\delta_n=0$ due to the translation invariance of the linear code $\mathcal{C}_1$ and uniformity of $X^n$.

\emph{Reconstruction}: After obtaining $\widehat{X}_{q}^n$ as in (\ref{eq:estimateXqn}), the secret-key is reconstructed in the GS model as 
\begin{align}
\widehat{S}&= \Dec_{\mathcal{C}}(\widehat{X}_{q}^n)
\end{align}
and in the CS model as 
\begin{align}
\widehat{S}'=\widehat{S}\xor(S\xor S')
\end{align}
both of which result in the same error probability.

\begin{remark}\label{rem:Maurerremark}
	We can improve the weak secrecy in (\ref{eq:secrecyleakconstruction}) and (\ref{eq:secrecyleakconstructionCS}) to strong secrecy, i.e., we can replace (\ref{eq:secrecyleakage_constraint}) with
	\begin{align}
	I(S;W)\leq \epsilon \qquad\qquad\qquad\text{(strong secrecy)}\label{eq:strongsecrecy_constraint}
	\end{align}
	by applying information reconciliation and privacy amplification steps to multiple blocks of identifier outputs as described in \cite{MaurerSecrecyFree}, e.g., by using multiple PUFs in a device for key agreement. 
\end{remark}

\begin{remark}
	We argue in Appendix~\ref{app:strongsecrecy} that there are code constructions that provide strong secrecy for general probability distributions $P_{XY}$ without additional information reconciliation and privacy amplification steps. 
\end{remark}

\subsection{Optimality of the Proposed Construction}
We prove that the above WZ-coding construction in combination with the proposed key agreement steps is optimal for the GS model.
\begin{theorem}\label{theo:GSoptimal}
	The key-leakage-storage region $\mathcal{R}_{\text{gs,bin}}$ in (\ref{eq:BSCRegionGS}) for the GS model is achieved by using the WZ-coding construction proposed above.
\end{theorem}
 
\begin{IEEEproof}
	By (\ref{eq:constraintonm1}) and (\ref{eq:constraintonm2}), we have 
	\begin{align}
	\frac{\log|\mathcal{W}|}{n}&\!=\!\frac{m_2}{n}\!=\!H_b(q*p_A) \!-\! H_b(q)+2\delta 
	\end{align} 
	so we set $R_w= H_b(q*p_A) - H_b(q)$. The secret key satisfies
	\begin{align}
	\frac{H(S)}{n}&\geq\frac{n-m_1-m_2}{n}-\delta=1- H_b(q*p_A)-2\delta
	\end{align}
	so we set $R_S= 1- H_b(q*p_A)$.	Furthermore, we have
	\begin{align}
	\frac{I(X^n;W)}{n}&\stackrel{(a)}{=} \frac{H(W)}{n}\leq\frac{\log|\mathcal{W}|}{n}=\frac{m_2}{n}\nonumber\\
	&=H_b(q*p_A) - H_b(q)+2\delta.
	\end{align}
	We thus set $R_\ell= H_b(q*p_A) - H_b(q)$, where $(a)$ follows because $X^n$ determines $W$.
\end{IEEEproof}

Combining Theorem~\ref{theo:GSoptimal} with the one-time padding idea discussed in Section~\ref{subsec:SKWZfirstconstruct}, we can show the optimality of the proposed code construction for the CS model also. 

\begin{remark}
	We show in Appendix~\ref{app:hiddenidentifiers} that the above WZ-coding construction is optimal also for hidden sources, i.e., the encoder observes a noisy measurement of the source rather than the source itself.
\end{remark} 

\section{Second WZ-coding Construction with Polar Codes}\label{sec:codeproposal}
Polar codes \cite{Arikan} have a low encoding/decoding complexity, asymptotic optimality for various problems, and good finite length performance if a list decoder is used. Furthermore, they have a structure that allows simple nested code design and they can be used for WZ-coding \cite{RudigerPolarExtended}. 

Polar codes rely on the \textit{channel polarization} phenomenon, where a channel is converted into polarized bit channels by a polar transform. This transform converts an input sequence $U^n$ with frozen and unfrozen bits to a codeword of the same length $n$. A polar decoder processes a noisy observation of the codeword together with the frozen bits to estimate ${U}^n$.

Let $\mathcal{C}(n,\mathcal{F},G^{|\mathcal{F}|})$ denote a polar code of length $n$, where $\mathcal{F}$ is the set of indices of the frozen bits and $G^{|\mathcal{F}|}$ is the sequence of frozen bits. In the following, we use the nested polar code construction proposed in \cite{RudigerPolarExtended}.

\subsection{Polar Code Construction for the GS Model}\label{subsec:polarcons}
We use two polar codes $\mathcal{C}_1(n,\mathcal{F}_1, V)$ and $\mathcal{C}(n,\mathcal{F}, \xoverline{V})$ with $\mathcal{F}=\mathcal{F}_1 \cup \mathcal{F}_w$ and $\xoverline{V}=[V, W]$, where $V$ has length $m_1$ and $W$ has length $m_2$ such that $m_1$ and $m_2$ satisfy (\ref{eq:constraintonm1}) and (\ref{eq:constraintonm2}). The indices in $\mathcal{F}_1$ represent frozen channels with assigned values $V$ for both codes and $\mathcal{C}$ has additional frozen channels with assigned values $W$ denoted by $\mathcal{F}_w$, i.e., the codes are nested. 

\begin{figure}
	\centering
	\begin{tikzpicture}[auto,>=latex', scale = 1.01, transform shape]
\centering
\scriptsize
\node[draw, rectangle,align=center, minimum height = 0.75cm, minimum width= 0.75 cm,rounded corners=.2cm](PUF){$P_X$};
\node[right of =PUF,node distance = 1cm](X){$X^n$};

\node[draw, right of=X,rectangle,align=center, node distance =2.25cm, minimum height = 0.75cm, minimum width= 1.8 cm, text width = 1.0cm,rounded corners=.2cm] (PYX) {$P_{Y|X}$};

\node[right of =PYX,node distance = 3cm](Y){};

\node[above of =Y,node distance = 1cm](Y2){$Y^n$};

\node[draw, above of=X,rectangle,align=center, node distance =2.5cm, minimum height = 0.9cm, minimum width= 1.4 cm, text width = 1.4cm] (VQ) {Polar Decoder $\mathcal{C}_1$};

\node[above of =VQ,node distance = 1cm](U){$U^n$};

\node[draw, above of=U,rectangle,align=center, node distance =1cm, minimum height = 0.9cm, minimum width= 1.4 cm, text width = 1.4cm] (SExt) {Helper Data and Key Extraction};

\node[above of =SExt,node distance = 1.25cm](S){$S$};

\node[right of =SExt,node distance = 2.5cm](W){$W$};

\node[draw, above of=Y,rectangle,align=center,node distance =2.5cm, minimum height = 0.9cm, minimum width= 1.4 cm, text width = 1.4cm] (Dec) {Polar Decoder $\mathcal{C}$};

\node[above of =Dec,node distance = 1cm](Uh){$\hat{U}^n$};

\node[draw, above of=Uh,rectangle,align=center, node distance =1cm, minimum height = 0.9cm, minimum width= 1.4 cm, text width = 1.4cm] (RExt) {Key Extraction};

\node[above of =RExt,node distance = 1.25cm](Sh){$\hat{S}$};

\node[left of =VQ,node distance = 1.36cm](V){$V$};
\node[left of =Dec,node distance = 1.65cm](VW){$W$};
\node[right of =Dec,node distance = 1.65cm](WV){$V$};

\node[draw, right of=VQ,rectangle,align=center, node distance =2.25cm, minimum height = 1cm, minimum width= 1.4 cm, text width = 1.4cm,dashed] (Enc) {Polar Transform};

\node[draw, above of=PYX,rectangle,align=center, node distance =1cm, minimum height = 0.75cm, minimum width= 1.8 cm, text width = 1.7cm,rounded corners=.2cm,dashed] (PXq) {BSC$(q*p_A)$};

\draw[decoration={markings,mark=at position 1 with {\arrow[scale=1.4]{latex}}},
postaction={decorate}, thick, shorten >=1.4pt] (PUF) -- (X); 
\draw[decoration={markings,mark=at position 1 with {\arrow[scale=1.4]{latex}}},
postaction={decorate}, thick, shorten >=1.4pt] (X) -- (PYX); 
\draw[decoration={markings,mark=at position 1 with {\arrow[scale=1.4]{latex}}},
postaction={decorate}, thick, shorten >=1.4pt] (PYX) -| (Y2);
\draw[decoration={markings,mark=at position 1 with {\arrow[scale=1.4]{latex}}},
postaction={decorate}, thick, shorten >=1.4pt] (X) -- (VQ);
\draw[decoration={markings,mark=at position 1 with {\arrow[scale=1.4]{latex}}},
postaction={decorate}, thick, shorten >=1.4pt] (VQ) -- (U);
\draw[decoration={markings,mark=at position 1 with {\arrow[scale=1.4]{latex}}},
postaction={decorate}, thick, shorten >=1.4pt] (U) -- (SExt);
\draw[decoration={markings,mark=at position 1 with {\arrow[scale=1.4]{latex}}},
postaction={decorate}, thick, shorten >=1.4pt] (SExt) -- (S);
\draw[decoration={markings,mark=at position 1 with {\arrow[scale=1.4]{latex}}},
postaction={decorate}, thick, shorten >=1.4pt] (SExt) -- (W);

\draw[decoration={markings,mark=at position 1 with {\arrow[scale=1.4]{latex}}},
postaction={decorate}, thick, shorten >=1.4pt] (Y2) -- (Dec);
\draw[decoration={markings,mark=at position 1 with {\arrow[scale=1.4]{latex}}},
postaction={decorate}, thick, shorten >=1.4pt] (Dec) -- (Uh);
\draw[decoration={markings,mark=at position 1 with {\arrow[scale=1.4]{latex}}},
postaction={decorate}, thick, shorten >=1.4pt] (Uh) -- (RExt);
\draw[decoration={markings,mark=at position 1 with {\arrow[scale=1.4]{latex}}},
postaction={decorate}, thick, shorten >=1.4pt] (RExt) -- (Sh);

\draw[decoration={markings,mark=at position 1 with {\arrow[scale=1.4]{latex}}},
postaction={decorate}, thick, shorten >=1.4pt] (V) -- (VQ);
\draw[decoration={markings,mark=at position 1 with {\arrow[scale=1.4]{latex}}},
postaction={decorate}, thick, shorten >=1.4pt] (VW) -- (Dec);
\draw[decoration={markings,mark=at position 1 with {\arrow[scale=1.4]{latex}}},
postaction={decorate}, thick, shorten >=1.4pt] (WV) -- (Dec);
\draw[decoration={markings,mark=at position 1 with {\arrow[scale=1.4]{latex}}},
postaction={decorate}, thick, shorten >=1.4pt] (W) -| (VW.north);

\draw[decoration={markings,mark=at position 1 with {\arrow[scale=1.4]{latex}}},
postaction={decorate}, thick, shorten >=1.4pt,dashed] (U) -| (Enc);

\draw[decoration={markings,mark=at position 1 with {\arrow[scale=1.4]{latex}}},
postaction={decorate}, thick, shorten >=1.4pt,dashed] (Enc) -- (PXq) node[midway] {$X_q^n$};
\draw[decoration={markings,mark=at position 1 with {\arrow[scale=1.4]{latex}}},
postaction={decorate}, thick, shorten >=1.4pt,dashed] (PXq) -- (Y2);

\node (redrect1) [thick,rectangle, draw, fit=(Dec)(RExt),inner sep=5pt,color=black,solid,rounded corners=.2cm] {};
\node (redrect2) [thick,rectangle, draw, fit=(VQ)(SExt),inner sep=5pt,color=black,solid,rounded corners=.2cm] {};
\small
\node[below of =X,node distance = 0.75cm](Enr){Enrollment};
\node[below of =Y,node distance = 0.75cm](Rec){Reconstruction};

\end{tikzpicture}
	\caption{Second WZ-coding construction for the GS model.}
	\label{fig:blockdig}
\end{figure}
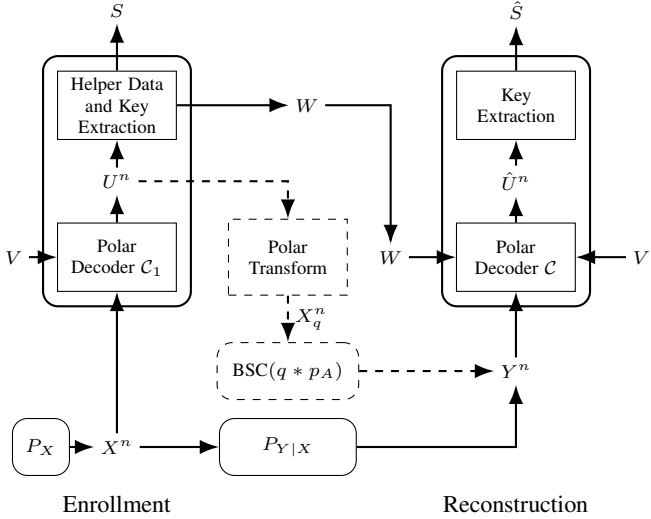

The code $\mathcal{C}_1$ serves as a VQ with a desired distortion $q$, and the code $\mathcal{C}$ serves as the error correcting code for a BSC($q*p_A$). The idea is to obtain $W$ during enrollment and store it as public helper data. For reconstruction, $W$ is used by the decoder to estimate the secret key $S$ of length $n-m_1-m_2$. Fig.~\ref{fig:blockdig} shows the block diagram of the proposed construction. In the following, suppose $V$ is the all-zero vector so that no additional storage is necessary. This choice has no effect on the average distortion $E[q]$ between $X^n$ and $X_q^n$ defined below; see \cite[Lemma 10]{RudigerPolarExtended}.

\textit{Enrollment}: The uniform binary sequence $X^n$ generated by a PUF during enrollment is treated as the noisy observation of a BSC$(q)$. $X^n$ is quantized by a polar decoder of $\mathcal{C}_1$. We extract from the decoder output $U^n$ the bits at indices $\mathcal{F}_w$ and store them as the helper data $W$. The bits at the indices $i\in \{1,2,\ldots,n\}\setminus \mathcal{F}$ are used as the secret key. Note that applying a polar transform to $U^n$ generates $X_q^n$, which is a distorted version  of $X^n$. The distortion between $X^n$ and $X_q^n$ is modeled as a BSC($q$) because the error sequence $E_{q}^n=X^n\xor X_q^n$ resembles an i.i.d. sequence $\sim\text{Bern}^n(q)$ when $n\rightarrow \infty$ \cite[Lemma 11]{RudigerPolarExtended}. 

\textit{Reconstruction}: During reconstruction, the polar decoder of $\mathcal{C}$ observes the binary sequence $Y^n$, which is a noisy measurement of $X^n$ through a BSC$(p_A)$. The frozen bits $\xoverline{V}=[V, W]$ at indices $\mathcal{F}$ are input to the polar decoder. The output $\widehat{U}^n$ of the polar decoder is the estimate of $U^n$ and contains the estimate $\widehat{S}$ of the secret key at the unfrozen indices of $\mathcal{C}$, i.e., $i\in \{1,2,\ldots,n\}\setminus \mathcal{F}$.

We next give a method to design practical nested polar codes for the GS model.

\textit{Construction of $\mathcal{C}$ and $\mathcal{C}_1$}: Since $\mathcal{C}\subseteq\mathcal{C}_1$ are nested codes, they must be constructed jointly. $\mathcal{F}$ and $\mathcal{F}_{1}$ should be selected such that the reliability and security constraints are satisfied. For a given secret key size $n-m_1-m_2$, block length $n$, crossover probability $p_A$, and target block-error probability $P_B=\Pr[S\ne\widehat{S}]$, we propose the following procedure.
\begin{enumerate}
	\item Construct a polar code of rate $(n\!-\!m_1\!-\!m_2)/n$ and use it as the code $\mathcal{C}$, i.e., define the set of frozen indices $\mathcal{F}$.
	\item Evaluate the error correction performance of $\mathcal{C}$ with a decoder for a BSC with a range of crossover probabilities to obtain the crossover probability $p_c$, resulting in a target block-error probability of $P_B$. Using $p_c=E[q]*p_A$, we obtain the target distortion $E[q]$ averaged over a large number of realizations of $X^n$.
	\item Find an $\mathcal{F}_1\subset \mathcal{F}$ that results in an average distortion of $E[q]$ with a minimum possible amount of helper data. Use $\mathcal{F}_1$ as the frozen set of $\mathcal{C}_1$. 
\end{enumerate}
Step 1 is a conventional polar code design task and step 2 is applied by Monte-Carlo simulations. For step 3, we start with  $\mathcal{F}_1^{'} = \mathcal{F}$ and compute the resulting average distortion $E[q']$ via Monte-Carlo simulations. If $E[q']$ is not less than $E[q]$, we remove elements from $\mathcal{F}_1^{'}$ according to the reliabilities of the polarized bit channels and repeat the procedure until we obtain the desired average distortion $E[q]$. 

We remark that the distortion level introduced by the VQ is an additional degree of freedom in choosing the code design parameters. For instance, different values of $P_B$ can be targeted with the same code by changing the distortion level. Alternatively, devices with different $p_A$ values can be supported by using the same code. This additional degree of freedom makes the proposed code design suitable for a wide range of applications. 

\subsection{Proposed Codes for the GS Model}
Consider, for instance, the GS model where $S$ is used in the advanced encryption standard (AES) with length 128, i.e., $\log|\mathcal{S}|=n-m_1-m_2=128$ bits. If we use PUFs in a field-programmable gate array (FPGA) as the randomness source, we must satisfy a block-error probability $P_B$ of at most $10^{-6}$ \cite{FPGAPUF}. Consider a BSC $P_{Y|X}$ with crossover probability $p_A=0.15$, which is a common value for SRAM PUFs under ideal environmental conditions \cite{maes2009soft} and for RO PUFs under varying environmental conditions \cite{bizimtemperature}. We design nested polar codes for these parameters to illustrate that we can achieve better key-leakage-storage rate tuples than previously proposed codes. 

\subsubsection*{Code 1} 
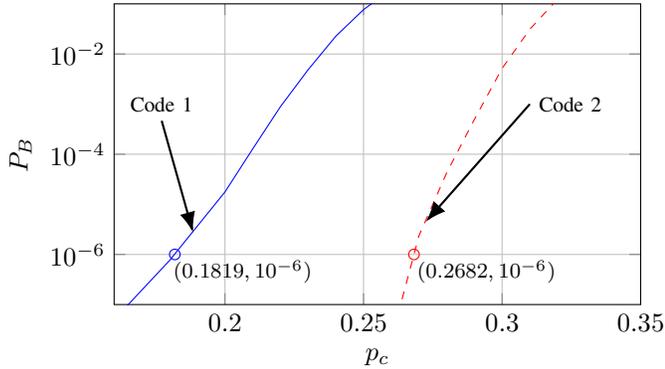
\begin{figure}[t]
%
%
\definecolor{mycolor1}{rgb}{0.00000,0,1}%
\definecolor{mycolor2}{rgb}{1.00000,0,0}%
\begin{tikzpicture}

\begin{axis}[%
width=7cm,
height=4cm,
at={(0.758in,0.481in)},
scale only axis,
xmin=0.16,
xmax=0.35,
xlabel={$p_c$},
xmajorgrids,
ymode=log,
ymin=1e-07,
ymax=1e-01,
yminorticks=true,
ylabel={$P_B$},
ymajorgrids,
yminorgrids,
axis background/.style={fill=white},
title style={font=\bfseries}
]
\addplot [color=mycolor1,solid,forget plot]
  table[row sep=crcr]{%
0.26	0.196908\\
0.25    0.076725\\
0.24	0.022648\\
0.23    0.004855\\
0.22	0.000887\\
0.21    0.0001275\\
0.2	1.76666666666667e-05\\
0.18	7.57575757575758e-07\\
0.16	5e-08\\
0.14	2e-08\\
};

\addplot [color=blue,only marks,mark=o,mark options={solid},forget plot]
table[row sep=crcr]{%
	0.1819	0.000001\\
};
\node (c1pc)[right, align=left, text=black]
at (axis cs:0.178,0.00000045) {\footnotesize$(0.1819,10^{-6})$};

\node (c1leg)[right, align=left, text=black]
at (axis cs:0.1625,0.001) {\footnotesize Code 1};

\draw[decoration={markings,mark=at position 1 with {\arrow[scale=1.5]{latex}}},
postaction={decorate}, thick, shorten >=1.4pt] (c1leg.south) -- ($(c1pc.west)+(10.5,1.8)$);


\addplot [color=mycolor2,dashed,forget plot]
table[row sep=crcr]{%
	0.35	0.855495\\
	0.34	0.622125\\
	0.33	0.333805\\
	0.32	0.12395\\
	0.31	0.031105\\
	0.3	0.005165\\
	0.29	0.000495\\
	0.28	4.25e-05\\
	0.27	2.35849056603774e-06\\
	0.26	2e-08\\
	0.25	0\\
	0.24	0\\
	0.23	0\\
	0.22	0\\
	0.21	0\\
	0.2	0\\
};

\addplot [color=red,only marks,mark=o,mark options={solid},forget plot]
table[row sep=crcr]{%
	0.2682	0.000001\\
};
\node (c2pc)[right, align=left, text=black]
at (axis cs:0.266,0.00000045) {\footnotesize$(0.2682,10^{-6})$}; 

\node (c2leg)[right, align=left, text=black]
at (axis cs:0.31,0.001) {\footnotesize Code 2};

\draw[decoration={markings,mark=at position 1 with {\arrow[scale=1.5]{latex}}},
postaction={decorate}, thick, shorten >=1.4pt] (c2leg.west) -- ($(c2pc.west)+(6.5,2.3)$);
\end{axis}
\end{tikzpicture}%
	\caption{Block-error probability of $\mathcal{C}$ over a BSC$(p_c)$ with an SCL decoder (list size 8) for Codes 1 and 2 of length $1024$ and $2048$, respectively.}
	\label{fig:n10242048comb}
\end{figure}

Consider $n=1024$ and recall that $n-m_1-m_2=128$, $P_B=10^{-6}$, and $p_A=0.15$. Polar successive cancellation list (SCL) decoders with list size $8$ are used as the VQ and channel decoder. We first design the code $\mathcal{C}$ of rate $128/1024$ and evaluate its performance with the SCL decoder for a BSC with a range of crossover probabilities, as shown in Fig.~\ref{fig:n10242048comb}. We observe a block-error probability of $10^{-6}$ at a crossover probability of $p_c=0.1819$. Since $p_A=0.15$, this corresponds to an average distortion of $E[q]=0.0456$, i.e., $E[q]*p_A=0.1819$.

Fig.~\ref{fig:q10242048comb} shows the average distortion $E[q]$ with respect to $n-m_1=n-|\mathcal{F}_1|$, obtained by Monte-Carlo simulations. We observe from Fig.~\ref{fig:q10242048comb} that the target average distortion is obtained at $n-m_1=778$ bits. Thus, $m_2=650$ bits of helper data suffice to obtain a block-error probability of $P_B=10^{-6}$ to reconstruct a $n-m_1-m_2=128$-bit secret key. 

We observe that the parameter $p_c$ is less than $p_A=0.15$ when we apply the procedure in Section~\ref{subsec:polarcons} to $n=512$ with the same $P_B$. Therefore, it is not possible to construct a code with our procedure for $n\leq 512$ since $q*p_A$ is an increasing function of $q$ for any $q\in [0, 0.5]$. Such a code construction for $n=512$ might be possible if one improves the code design and the decoder.

\subsubsection*{Code 2} 
\begin{figure}[!t]
	\input{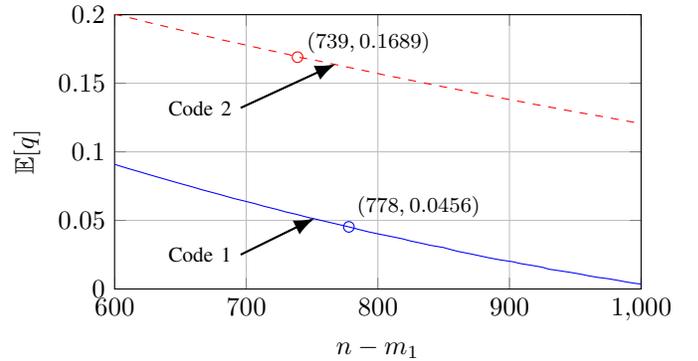}
	\caption{Average distortion $E[q]$ with respect to $n-m_1$ with an SCL decoder (list size 8) for Codes 1 and 2 of length $1024$ and $2048$, respectively.}\label{fig:q10242048comb}
\end{figure}

Consider the same parameters as in Code 1, except $n=2048$. We apply the same steps as above and plot the performance of an SCL decoder for a BSC with a range of crossover probabilities in Fig.~\ref{fig:n10242048comb}. A crossover probability of $p_c=0.2682$ is required to obtain a block-error probability of $10^{-6}$, which gives an average distortion of $E[q]=0.1689$. As depicted in Fig.~\ref{fig:q10242048comb}, we achieve the target average distortion with $n-m_1=739$ bits so that helper data of length $611$ bits is required to satisfy $P_B=10^{-6}$ for a secret key of length $128$ bits. 

\begin{remark}
	Our assumptions on the channel statistics are not necessarily satisfied for the model depicted in Fig.~\ref{fig:blockdig} for finite $n$ since, e.g., the channel $P_{X^n|X^n_q}$ is not $\sim\text{Bern}^n(q)$. However, our code designs and analysis are based on simulations made over a large number of possible inputs at fixed lengths, which allows us to give reliability guarantees to a set of input realizations. The results of such guarantees are given below. 
\end{remark}

The error probability $P_B$ is calculated as an average over a large number of PUF realizations, i.e., over a large number of PUF devices with the same circuit design. To satisfy the block-error requirement for each PUF realization, one could consider using the maximum distortion instead of $E[q]$ as a metric in step 3 in Section~\ref{subsec:polarcons}. This would increase the amount of helper data. We can guarantee a block-error probability of at most $10^{-6}$ for $99.99\%$ of all realizations $x^n$ of $X^n$ by adding $32$ bits to the helper data for Code 1 and $33$ bits for Code 2. The numbers of extra helper data bits required are small since the variance of the distortion $q$ over all PUF realizations is small for the blocklengths considered. For comparisons, we use the helper data sizes required to guarantee $P_B=10^{-6}$ for $99.99\%$ of all PUF realizations.
\begin{figure*}[t] 
	\centering
	\newlength\figureheight
	\newlength\figurewidth
	\setlength\figureheight{5.75cm}
	\setlength\figurewidth{16.4cm}
	\input{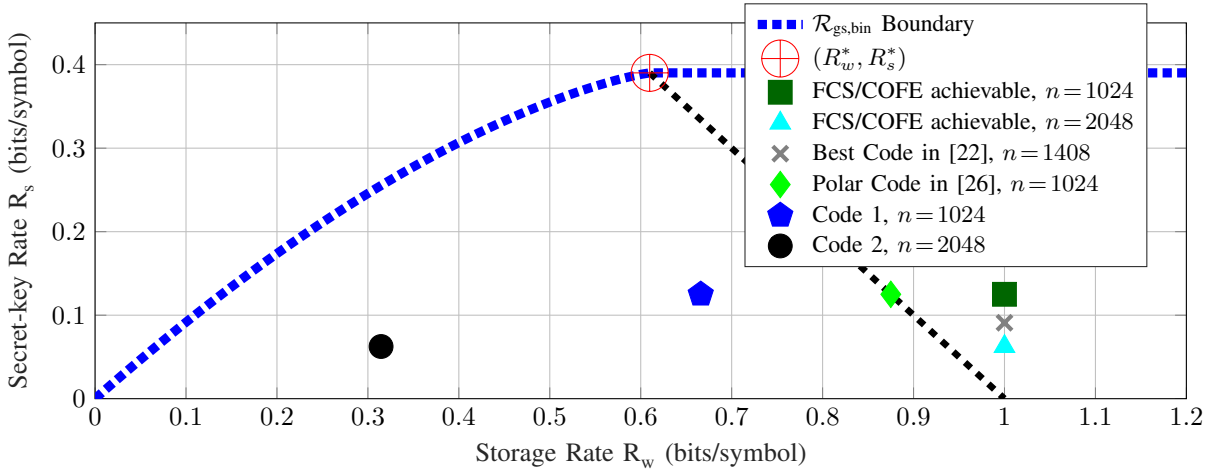}
	\caption{Storage-key rates for the GS model with $p_A=0.15$. The $(R_w^*,R_s^*)$ point is the best possible point achieved by SW-coding constructions, which lies on the dashed line representing $R_w+R_s = H(X)$. The block-error probability satisfies $P_B \leq 10^{-6}$ and the key length is 128 bits for all code points.}
	\label{fig:codecomparisons}
\end{figure*} 
\vspace{-0.45cm}
\subsection{Code Comparisons and Discussions}
We show in Fig.~\ref{fig:codecomparisons} the storage-key $(R_w,R_s)$ projection of the boundary points of the rate region $\mathcal{R}_{\text{gs,bin}}$ in (\ref{eq:BSCRegionGS}) for $p_A=0.15$. Furthermore, we show the point with the maximum secret-key rate $R_s^*$ and the minimum storage rate $R_w^*$ to achieve $R_s^*$. For the FCS and COFE, we use the random coding union bound \cite[Thm. 16]{Polyansky} to confirm that the plotted rate pairs are achievable for a secret-key length of $128$ bits, a block-error probability of $P_B=10^{-6}$, and blocklengths of $n=1024$ and $n=2048$. For the same key length and block-error probability we also plot the best code given in \cite{maes2009soft}, which is a concatenation of an inner $(4,1)$ repetition code and an outer $(32,16)$ Reed-Muller (RM) code. In \cite{maes2009soft}, the COFE is used for a block of biased SRAM PUFs. These rate pairs are shown in Fig.~\ref{fig:codecomparisons} to the right of the dashed line representing $R_w+R_s=1$. Similarly, the rate pairs achieved by the polar code design in \cite{IgnaPolar} and Codes 1 and 2 are shown in Fig.~\ref{fig:codecomparisons}. 

The storage rates of the FCS and COFE are 1 bit/symbol, which is suboptimal as discussed in Section~\ref{sec:CompareMethods}. The polar code construction in \cite{IgnaPolar} achieves a rate point with $R_s+R_w =1$ bit/symbol, which is expected since this is a SW-coding construction. The polar code construction improves on the rate pairs achieved by the FCS and COFE in terms of the key vs. storage ratio. Note that the FCS, COFE, and the previous polar code construction achieve the privacy-leakage rate $R_\ell= 1-R_s$ bits/symbol; see \cite[Eq. (21)]{IgnaFuzzy}.
	
We achieve the key-leakage-storage rates of approximately $(0.125, 0.666, 0.666)$ bits/symbol by Code 1 and $(0.063, 0.315, 0.315)$ bits/symbol by Code 2, projections of which are depicted in Fig.~\ref{fig:codecomparisons}. These rates are significantly better than the best rate tuple $(0.125, 0.875, 0.875)$ bits/symbol in the literature, i.e., the polar code construction in \cite{IgnaPolar}, for the same constraints and without any private key assumption. We increase the key vs. storage rate ratio $R_s/R_w$ from $0.188$ for Code 1 to $0.199$ for Code 2, which suggests to increase the blocklength to obtain better ratios. Furthermore, Code 2 achieves privacy-leakage and storage rates that cannot be achieved by existing methods without applying \textit{time sharing} (see, e.g., \cite[Section 4.4]{Elgamalbook}). This is because Code 2 achieves privacy-leakage and storage rates of $0.315$ bits/symbol that are significantly less than the minimum privacy-leakage and storage rates $R^*_w=R^*_\ell=H_b(p_A)\approxeq0.610$ bits/symbol that can be asymptotically achieved by existing methods at the maximum secret-key rate $R_s^*\approxeq 0.390$ bits/symbol. 

We use the sphere packing bound \cite[Eq. (5.8.19)]{gallagerbook} to upper bound the key vs. storage rate ratio that can be achieved by SW-coding constructions for the maximum secret-key rate point. Consider $p_A=0.15$, $n=1024$, and $P_B=10^{-6}$, for which the sphere packing bound requires that the rate of the code $\mathcal{C}$ satisfies $R_\mathcal{C} \leq 0.273$. If we assume that the key rate is given by its maximal value $R_s = R_\mathcal{C}$ and the storage rate is given by its minimal value $R_w = 1 - R_\mathcal{C}$, then we arrive at $R_s/R_w \le 0.375$. A similar calculation for $n=2048$ yields $R_s/R_w \le 0.437$. These results indicate that there are still gaps between the maximum key vs. storage rate ratios achieved by WZ-coding constructions, which might achieve higher ratios than SW-coding constructions, and the ratios achieved by Codes 1 and 2. The gaps can be reduced by using, e.g., larger list sizes at the decoder, but this is not desired for IoT applications that require low hardware complexity. For other PUF applications, codes that satisfy $P_B\leq10^{-9}$ should be designed \cite{OurEntropy}, for which either laborious decoder simulations or analytical block-error probability bounds seem to be required.
 
\section{Conclusion}
We argued that a WZ-coding construction based on random linear codes is asymptotically optimal for the GS and CS models with uniform binary sources with decoder measurements through a BSC. These source and channel models are standard models for RO PUFs and SRAM PUFs. We implemented a second WZ-coding construction with nested polar codes that achieve better rate tuples than existing methods, and one of our codes achieves a rate tuple that cannot be achieved by existing methods without time sharing. Gaps to the maximum key vs. storage rate ratios were illustrated. 
\section*{Acknowledgment}
O. G\"unl\"u thanks Amin Gohari and Navin Kashyap for their insightful comments, and also Matthieu Bloch for his help and useful suggestions that significantly improved this work.   

\appendices


\section{Strong Secrecy}\label{app:strongsecrecy}
\begin{theorem}\label{theo:strongsecrecy}
	For the GS model (or CS model), given any $\epsilon>0$, there exist some $n\geq 1$, an encoder, and a decoder that achieve the key-leakage-storage region $\mathcal{R}_{\text{gs}}$ (or $\mathcal{R}_{\text{cs}}$) and that satisfy the strong-secrecy constraint in (\ref{eq:strongsecrecy_constraint}). 
\end{theorem}

We give proof sketches for Theorem~\ref{theo:strongsecrecy} for the GS model by using two approaches; the first uses output statistics of random binning (OSRB) \cite{OSRBAmin} and the second uses resolvability \cite{GerhardDivergence} and a likelihood encoder \cite{LikelihoodEncoder}. The proofs for the CS model follow by applying a one-time pad step, as in Section~\ref{subsec:SKWZfirstconstruct}.

\begin{IEEEproof}[Proof Sketch 1]We first give a random binning based proof by following the steps in \cite{OSRBAmin}. Fix a $\displaystyle P_{U|X}$ and let $(U^n,X^n,Y^n)$ be i.i.d. according to $P_{U|X}P_XP_{Y|X}$. For each $u^n$, assign three random bin indices $S\in[1:2^{nR_s}]$, $W\in[1:2^{nR_w}]$, and $C\in[1:2^{nR_c}]$, which represent, respectively, the secret key, helper data, and randomness shared by encoder, decoder, and eavesdropper (similar to $W$). 
	
We use a SW decoder to estimate $\widehat{U}^n$ from $(C,W,Y^n)$, which satisfies (\ref{eq:reliability_constraint}) if (see \cite[Lemma 1]{OSRBAmin})
\begin{align}
	R_c+R_w> H(U|Y). \label{eq:slepianwolfdecoder}
\end{align}
	
We further have that $(S,W,C)$ are almost mutually independent and uniform so that (\ref{eq:uniformity_constraint}) and (\ref{eq:strongsecrecy_constraint}) are satisfied if we have (see \cite[Theorem 1]{OSRBAmin})
\begin{align}
	R_s+R_w+R_c< H(U).\label{eq:independenceofindices}
\end{align}
Similarly, the shared randomness $C$ is almost independent of $X^n$, suggesting that it is almost independent of $Y^n$ also, if 
\begin{align}
	R_c<H(U|X).\label{eq:independenceofcode}
\end{align} 
	
Applying Fourier-Motzkin elimination \cite[Section 12.2]{FourierMotzkin} to (\ref{eq:slepianwolfdecoder})-(\ref{eq:independenceofcode}), we can show that there exists a binning with a fixed value of $C$ and it achieves all rate tuples $(R_s,R_{\ell},R_w)$ in the key-leakage-storage region $\mathcal{R}_{\text{gs}}$ with strong secrecy. 
\end{IEEEproof}

\begin{IEEEproof}[Proof Sketch 2]We next sketch a random coding based proof by following the steps in \cite{LikelihoodEncoder} and \cite[Section 1.6.2]{BlochLectureNotes2018}. Consider the channel coding problem where $S\in[1:2^{nR_s}]$ and $W\in[1:2^{nR_w}]$ are uniform and independent inputs of an encoder $\Enc(\cdot)$ with the output codeword $U^n$ that passes through a channel $P_{X|U}$ to obtain $X^n$, which further passes through the channel $P_{Y|X}$ to obtain $Y^n$. Applying the resolvability result from \cite[Theorem 1]{GerhardDivergence}, one can simulate $X^n\sim \prod_{i=1}^nP_X(x_i)$ if 
	\begin{align}
	R_s+R_w>I(U;X).
	\end{align}
	Furthermore, one can reliably estimate $\widehat{U}^n$ from $(W,Y^n)$ if
	\begin{align}
	R_s< I(U;Y).
	\end{align} 
	Note that this channel coding problem defines a joint probability distribution 
	\begin{align}
	&\widetilde{P}_{SWX^nY^n}(s,w,x^n,y^n)\nonumber\\
	&\!=\! Q^{\text{Unif}}_S(s)Q^{\text{Unif}}_W(w)\mathds{1}\{x^n\!=\!\Enc(w,s)\}\prod_{i=1}^nP_{Y|X}(y_i|x_i) \label{eq:alliedprobability}
	\end{align}
	where $Q^{\text{Unif}}_S$ and $Q^{\text{Unif}}_W$ are uniform probability distributions over the sets, respectively, $[1:2^{nR_s}]$ and $[1:2^{nR_w}]$, and $\mathds{1}\{\cdot\}$ is the indicator function.
	
	However, for the original problem, we should invert the random coding and use a stochastic encoder according to the conditional probability distribution $\widetilde{P}_{SW|X^n}$ obtained from (\ref{eq:alliedprobability}), which induces a joint distribution
	\begin{align}
	&P_{SWX^nY^n}(s,w,x^n,y^n)\nonumber\\
	&\qquad\quad = \widetilde{P}_{SW|X^n}(s,w|x^n) \prod_{i=i}^nP_X(x_i)P_{Y|X}(y_i|x_i).
	\end{align}
	It follows from the above channel coding problem that (\ref{eq:reliability_constraint}), (\ref{eq:uniformity_constraint}), (\ref{eq:storage_constraint}), and (\ref{eq:strongsecrecy_constraint}) are satisfied. We can show that there exist some $n\geq 1$, an encoder, and a decoder that achieve all rate tuples $(R_s,R_{\ell},R_w)$ in the key-leakage-storage region $\mathcal{R}_{\text{gs}}$ with strong secrecy.
\end{IEEEproof}

\begin{remark}
	A random linear code (RLC) construction for binary input channels \cite{GerhardRLC} can provide resolvability for the bins $(S,W)$ with strong secrecy. A binary $U$ is optimal for the rate regions $\mathcal{R}_{\text{gs}}$ and $\mathcal{R}_{\text{cs}}$ if, e.g., $P_{Y|X}$ can be decomposed into a mixture of BSCs \cite[Theorem 3]{bizimTIFSMultipleMeasurement}.
\end{remark}

\begin{remark}\label{rem:RemiPolarNotPractical}
	In \cite[Theorem 10]{RemiPolarConstruction}, a polar code construction based on OSRB is shown to be optimal for the GS model with strong secrecy. This construction requires chains of identifier outputs, each of which has size $n$, and a secret seed shared between the encoder and decoder. For a finite blocklength, the secret seed size can be large and it can be seen as a ``private" key, which is not realistic as discussed in Section~\ref{sec:CompareMethods} for some code constructions in \cite{IgnaPolar}. Furthermore, the constructions used in Proofs 1 and 2 of Theorem~\ref{theo:strongsecrecy} are stochastic and such code constructions do not seem to be practical.
\end{remark}


\section{Extensions to Hidden Sources with Multiple Decoder Measurements}\label{app:hiddenidentifiers}
The GS and CS models in Fig.~\ref{fig:problemsetup} are extended in \cite{bizimTIFSMultipleMeasurement} by having the encoder measure a noisy version $\widetilde{X}^n$ of a hidden, or remote, identifier source $X^n$. The encoder generates or embeds a secret key and sends a public message $W$ or $W'$ to the decoder. The decoder observes another noisy measurement $Y^n$ of the source and estimates the secret key. The key-leakage-storage regions that satisfy (\ref{eq:reliability_constraint})-(\ref{eq:leakage_constraint}) for the GS and CS models with a hidden source are given in the following theorem.

\begin{theorem}[\hspace{1sp}\cite{bizimTIFSMultipleMeasurement}]\label{theo:regionsforhiddensource}
	The key-leakage-storage regions for the GS and CS models with a hidden source, respectively, are 
	\begin{align}
	\widetilde{\mathcal{R}}_{\text{gs}}\! =\! &\bigcup_{P_{U|\widetilde{X}}}\!\Big\{\left(R_s,R_\ell,R_w\right)\!\colon\!\nonumber\\
	&0\leq R_s\leq I(U;Y),\nonumber\\
	&R_\ell\geq I(U;X) - I(U;Y),\nonumber\\
	&R_w\geq I(U;\widetilde{X})- I(U;Y)\Big\}\label{eq:rateregionhiddengenerated},
	\end{align}
	\vspace*{-0.2cm}
	\begin{align}
	\widetilde{\mathcal{R}}_{\text{cs}}\! =\! &\bigcup_{P_{U|\widetilde{X}}}\!\Big\{\left(R_s,R_\ell,R_w\right)\!\colon\! \nonumber\\
	&0\leq R_s\leq I(U;Y),\nonumber\\
	&R_\ell\geq I(U;X) - I(U;Y),\nonumber\\
	&R_w\geq I(U;\widetilde{X})\Big\}\label{eq:rateregionhiddenchosen}
	\end{align}
	where $U-\widetilde{X}-X-Y$ forms a Markov chain. These regions are convex sets. The alphabet $\mathcal{U}$ of the auxiliary random variable $U$ can be limited to have size $\displaystyle |\mathcal{U}|\!\leq\!|\mathcal{\widetilde{X}}|+2$. 
\end{theorem}

Suppose next that the encoder measures a binary hidden source $X^n$ through a channel $P_{\widetilde{X}|X}$ such that the inverse channel $P_{X|\widetilde{X}}$ is a BSC, and the decoder measures the source through a channel $\displaystyle P_{Y|X}$ that is a BSC.  

\begin{theorem}[\hspace{1sp}\cite{bizimTIFSMultipleMeasurement}]\label{theo:OptimalityofBSCUtoXtilde}
	Assume $P_{X|\widetilde{X}}$ is a BSC and $\displaystyle P_{Y|X}$ is a binary-input symmetric memoryless channel. The boundary points of $\displaystyle \widetilde{\mathcal{R}}_{\text{gs}}$ and $\displaystyle \widetilde{\mathcal{R}}_{\text{cs}}$ are achieved by channels $\displaystyle P_{\widetilde{X}|U}$ that are BSCs. 
\end{theorem}

We next argue the optimality of the first WZ-coding construction given in Section~\ref{sec:firstWZconstruction} for the GS and CS models with the hidden source model considered above.

\begin{theorem}
	The WZ-coding construction given in Section~\ref{sec:firstWZconstruction} achieves the regions $\displaystyle \widetilde{\mathcal{R}}_{\text{gs}}$ and $\displaystyle \widetilde{\mathcal{R}}_{\text{cs}}$ for a uniform source $X^n$, an inverse channel $P_{X|\widetilde{X}}$ that is a BSC, and a decoder-measurement channel $\displaystyle P_{Y|X}$ that is also a BSC.
\end{theorem}

\begin{IEEEproof}
	We modify the WZ-coding construction in Section~\ref{sec:firstWZconstruction} by defining the new error sequence 
	\begin{align}
	\widetilde{E}_q^n = \widetilde{X}^n\xor \widetilde{X}_{q}^n\label{eq:newerrorsequence}
	\end{align}
	which resembles an i.i.d. sequence $\sim \text{Bern}^n(q)$ for some $q\in[0,0.5]$ when $\widetilde{X}_q^n$ is the closest codeword of $\mathcal{C}_1$ to $\widetilde{X}^n$ in Hamming distance and when $n\rightarrow\infty$. The new error sequence represents the BSC $\displaystyle P_{\widetilde{X}|U}$ since the new common randomness $\widetilde{X}_q^n$ asymptotically represents the auxiliary random variable $U^n$. Therefore, we asymptotically obtain the memoryless channel $\displaystyle P_{\widetilde{X}|U}\sim \text{BSC}(q)$. It follows from Theorem~\ref{theo:OptimalityofBSCUtoXtilde} that applying the modified code construction and taking a union of the rate tuples achieved over all $q\in[0, 0.5]$, we can achieve the boundary points of $\displaystyle \widetilde{\mathcal{R}}_{\text{gs}}$ and $\displaystyle \widetilde{\mathcal{R}}_{\text{cs}}$. 
\end{IEEEproof}

\begin{remark}
	Applying additional information reconciliation and privacy amplification steps to multiple identifier blocks, as in Remark~\ref{rem:Maurerremark}, provides strong secrecy also for hidden sources. Alternatively, random binning and random coding based approaches can be applied, as in Theorem~\ref{theo:strongsecrecy}, to show that there exist code constructions that provide strong secrecy. 
\end{remark}

\ifCLASSOPTIONcaptionsoff
\newpage
\fi

\bibliographystyle{IEEEtran}
\bibliography{IEEEabrv,references}

\end{document}